\documentclass[a4paper,article]{jpconf}
\usepackage{graphicx}
\usepackage{float}
\begin{document}
\title{Signatures
of Flux Tube Fragmentation and Strangeness Correlations in $pp$ Collisions }

\author{Cheuk-Yin Wong } 

\address{Physics Division, Oak Ridge National Laboratory, Oak Ridge, TN 37831, USA} 

\ead{wongc@ornl.gov}

\begin{abstract}

In the fragmentation of a color flux tube in high-energy $pp$
collisions or $e^+$-$e^-$ annihilations, the production of $q$-$\bar
q$ pairs along a color flux tube precedes the fragmentation of the
tube.  The local conservation laws in the production of these
$q$-$\bar q$ pairs will lead to the correlations of adjacently
produced hadrons.  As a consequence, the fragmentation of a flux tube
will yield a many-hadron correlation in the form of a chain of hadrons
ordered in rapidity, with adjacent hadrons correlated in charges,
flavor contents, and azimuthal angles.  It will also lead to a
two-hadron angular correlation between two hadrons with opposite
charges or strangeness that is suppressed at $\Delta \phi\sim 0$ but
enhanced at $\Delta \phi\sim \pi$, within a rapidity window $\Delta
y $$\sim$$1/(dN/dy)$.

\end{abstract}

\vspace*{-1.0cm}

\section{Introduction}

\hspace*{0.3cm} The reaction mechanisms in $pp$ collisions provide
insights in $AA$ collision at high energies.  The importance of each
reaction mechanism depends on the collision energy and the $p_T$
domain.  The flux-tube fragmentation process \cite{Art74,And79} is
expected to dominate at low $p_T$ and low $\sqrt{s_{pp}}$, whereas the
hard-scattering process \cite{Bla74,Ang78} at high $p_{T}$ and high
$\sqrt{s_{pp}}$.  The two mechanisms cross-over at a certain
transverse momentum $p_{Tb}(\sqrt{s_{pp}})$ that is a function of the
collision energy, $\sqrt{s_{pp}}$.  There are in addition other
mechanisms such as the recombination of partons \cite{Hwa80} and the
production of resonances.

It is desirable to establish, even if approximately, the boundary
function $p_{Tb}(\sqrt{s_{pp}})$ that separates the region of flux-tube
fragmentation dominance from the region of hard-scattering dominance.
We need signatures for these two mechanisms to facilitate such a
separation.

The signature for the hard scattering process is well known.  It is
given as a two-hadron $\Delta \phi-$$\Delta \eta$ angular correlation
which shows a peak at $ (\Delta \phi,\Delta \eta)$$\sim$0 and a ridge
along $\Delta \eta$ at $ \Delta \phi \sim \pi$.  The peak at
$\Delta\phi$$\sim$0 arises from the fragmentation of the jet
associated with the trigger and the ridge at $\Delta\phi$$\sim$$\pi$
arises from the fragmentation of the other jet associated with
colliding partons with unbalanced longitudinal momenta.

However, the signature for flux tube fragmentation has not been well
studied.  We would like to present here two signatures for flux-tube
fragmentation \cite{Won15,Won15a}.

\section{Many-hadron signatures of Flux Tube Fragmentation}

\hspace*{0.3cm} In the semi-classical description of the flux-tube
fragmentation process \cite{Art74,And79} for hadron production, the
production of quark-antiquark pairs along a color flux tube precedes
the fragmentation of the tube.  The production of these
quark-antiquark pairs must however obey conservation laws at the local
production points.  As a consequence, the produced $q$-$\bar q$ pairs
will lead to correlations of adjacently produced hadrons, and the
hadrons are ordered according to their rapidities along the tube.  The
rapidity-space-time ordering and the local conservation laws will
yield a many-body correlation of the hadrons in charge, flavor, and
momentum, which may provide vital information on space-time dynamics
of quarks and hadrons in the flux-tube fragmentation process.

As an example, we examine the fragmentation of a flux tube with an
invariant mass of 8.65 GeV, corresponding to the average invariant
mass of one of the two flux tubes in a $pp$ collision at 17.3 GeV,
with the flux tube formed by a quark of one proton with the diquark of
the other proton.  We carry out a Monte Carlo generation of hadrons in
flux-tube fragmentation using the PYTHIA 6.4 program \cite{Sjo06}.  An
example of the produced hadrons involving the production of a pair of strange
hadrons is shown in Fig. 1, with the production of 5 hadrons
listed in Table I.

\begin{figure}[H]
\centering
\label{fig1}
\includegraphics[width = 400 pt, height = 185 pt]{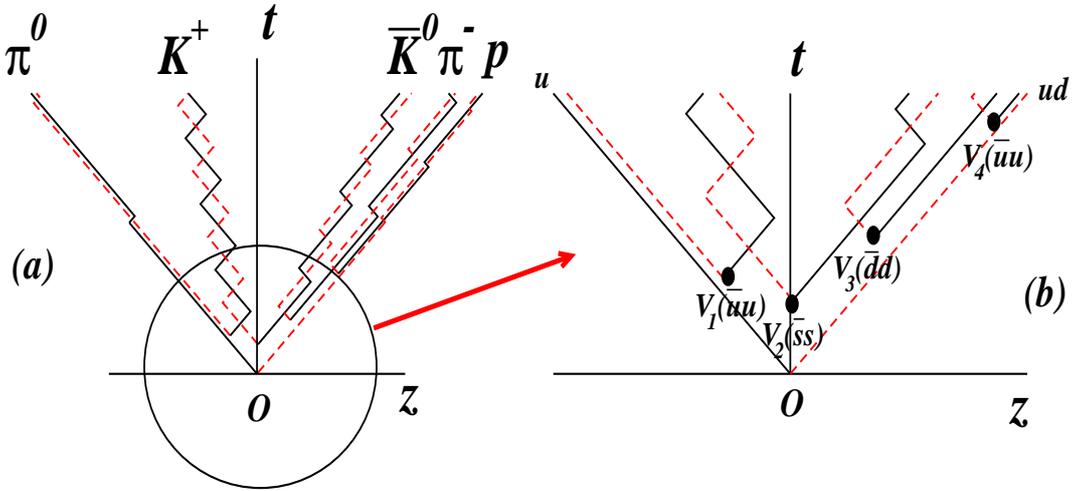} 
\caption{\label{etaphi} 
Fig. 1(a) shows an example of the space-time diagram in the
fragmentation of a $u$-$(ud)$ flux tube with an invariant mass of
$\sqrt{s}$=8.65 GeV, obtained with PYTHIA 6.4 \cite{Sjo06}.  Quark
lines are shown as solid lines, and antiquark (or diquark) lines as
dashed lines. Fig. 1(b) is an expanded view of the pair-production
vertices $V_i(\bar q_i  q_i)$ of Fig. 1(a).}
\end{figure}

\begin{table}[H]
\vspace*{-0.8cm}
  \caption{
 An example of primary hadrons $i$, their rapidities $y_i$, their
 azimuthal angles $\phi_i$ and their constituents $q_i$-$\bar q_i$
 produced in the fragmentation of the $u$-$(ud)$ flux tube at an
 energy  of $\sqrt{s}$=8.65 GeV obtained with PYTHIA 6.4 \cite{Sjo06}. }

\vspace*{0.3cm}
\centering

\begin{tabular}{|c | c c c c c|}
\hline
$i$  & 1 & 2   & 3 &  4 & 5 \\
particle
  & $\pi^0$ &  $K^+$ & $ \bar K^0$ & $\pi^-$ & $p$    \\ \hline
\noalign{\vskip 0.06cm}  
$ q_i$-$\bar  q_i$ 
& $u$-$\bar u$ &  $u$-$\bar s$ & $s$-$\bar d$ &   $ d$ -$\bar u$  & $u$ -$(ud)$
   \\
 $y_i$  
  &   -1.55  &  - 1.15&  - 0.75 & 0.27&  1.78 \\
$\phi_i$      &  1.00  & -2.01  &  1.44  & -2.43   &   0.17  \\

$\phi_{i}$$-$$\phi_{i+1}$  &\multicolumn{5}{c|}{ \hspace*{-0.4cm}~~-3.01   ~~ ~3.46   ~ ~-3.87   ~ ~  ~2.60   } \\

\hline
\end{tabular}
\end{table}
\noindent
In Table 1, the row of $q_i$-$\bar q_i$ shows that upon ordering the
hadrons according to their repidities $y_i$ as in a chain, the flavors of
the constituent antiquark $\bar q_i$ and the flavors of the neighboring
constituent quark $q_{i+1}$ are correlated along the chain, on an
event-by event basis.  The row of $\phi_i$-$\phi_{i+1}$ of neighboring
hadrons in Table I indicates that neighboring pairs of hadrons are
azimuthally correlated, approximately in a back-to-back manner.  The
many-hadron signature requires the identification of all hadrons
detected in the events. Predicted signature of this kind is yet to be
observed.

\section{Two-hadron angular correlation signature
of flux-tube fragmentation}

\hspace*{0.3cm} Another signature of the flux tube fragmentation
utilizes the two-hadron angular correlations arising from the
productions of $q$-$\bar q$ pairs.  Because of local conservation
laws, the production of $q$-$\bar q$ pairs will lead to correlations
of adjacently produced hadrons.  Adjacently produced hadrons however
can be signaled by their rapidity difference $\Delta y$ falling within
the window of $|\Delta y | $$\sim$$ 1/(dN/dy)$, on account of the
space-time-rapidity ordering of produced mesons in a flux-tube
fragmentation.  Therefore, the local conservation laws of momentum,
charge, and flavor will lead to a suppression of the angular
correlation function $dN/(d\Delta \phi\, d\Delta y)$ for two hadrons
with opposite charges or strangeness at $(\Delta \phi, \Delta y)$$
\sim$0, but an enhanced correlation on the back-to-back, away side at
$\Delta \phi$$\sim$$ \pi$, within the window of $|\Delta y |$$\sim$$
1/(dN/dy)$.  When we approximate the rapidity $y$ as the
pseudorapidity $\eta$, the two-hadron angular correlations can be used
as signatures for the fragmentation of a color flux tube as shown in
Fig 2.

\vspace*{-3.0cm}
\hspace{5.0cm}
\begin{figure}[H]
\hspace{-0.0cm}
\includegraphics[width = 250 pt, height = 250 pt]{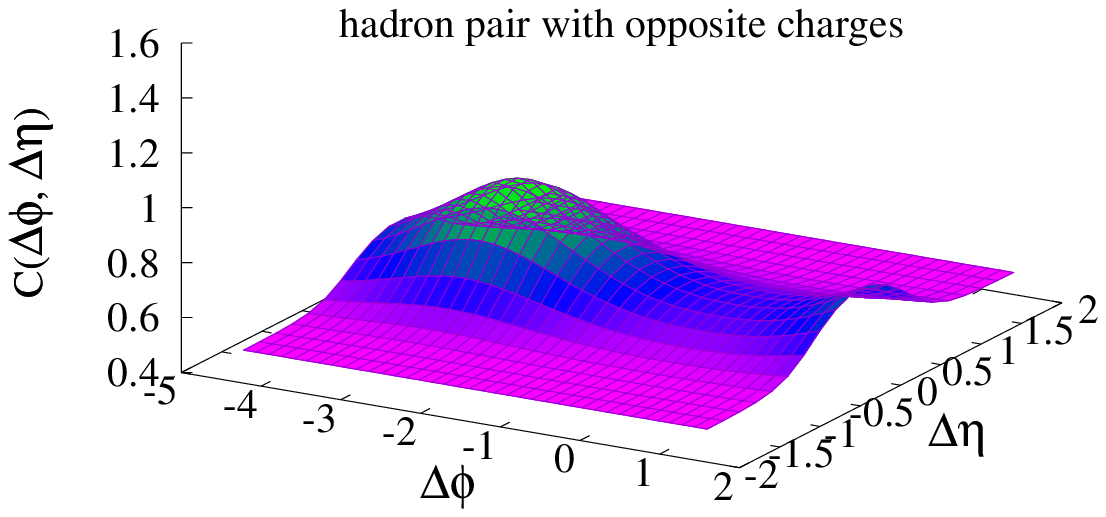}
\end{figure}
\vspace*{-1.5cm}
\hspace*{3.7cm}
{\Large \bf (a) } 

\vspace*{-6.2cm}
\begin{figure}[H]
\hspace*{8.5cm}
\includegraphics*[width = 250 pt, height = 420 pt]{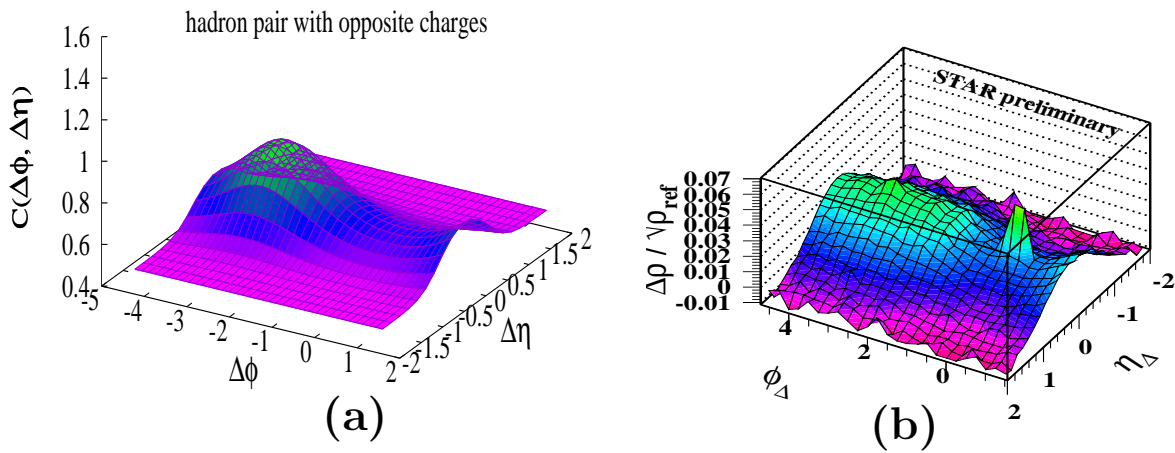}
\vspace*{-10.1cm}
\hspace*{4.5cm}

\hspace*{11.4cm}
{\Large \bf (b) } 
\caption{
Fig. 2(a) gives the theoretical correlations for two two hadrons with
unlike charges \cite{Won15} and Fig 2(b) gives  the experimental
correlation data  for two hadrons with unlike charges and $p_T$$<$0.5 GeV
obtained by the STAR Collaboration for $pp$ collisions at
$\sqrt{s_{pp}}$=200 GeV \cite{Por05}.  The sharp peak at $(\Delta
\phi, \Delta\eta)\sim 0$ in Fig. 2(b) is an experimental artifact.  }
\end{figure}

Comparison of theoretical angular correlations for two hadrons with
unlike charges in Fig.\ 2(a) \cite{Won15} with experimental data from
the STAR Collaboration in Fig.\ 2(b) \cite{Por05} indicates that in
high-energy $pp$ collisions at $\sqrt{s_{pp}}=$200 GeV, the production
of unlike-charge hadron pairs in the region of $p_T$$<$0.5 GeV/c are
qualitatively consistent with the flux-tube fragmentation mechanism.
However, the correlations for two hadrons with unlike-charges in the
region with $p_T$$>$0.5 GeV/c exhibit a completely different pattern.
Namely, they shows a peak at $(\Delta \phi, \Delta \eta)\sim$ 0, and a
ridge along $\Delta\eta$ at $\Delta\phi\sim \pi$, which is a signature
of the hard scattering process \cite{Por05}.  This indicates that for
$pp$ collisions energy at $\sqrt{s_{pp}}=200$ GeV, the boundary
between the flux-tube fragmentation process and the hard-scattering
process is $p_{Tb}=0.5$ GeV/c.  At the LHC energy of $\sqrt{s}_{pp}=$5
TeV, the minimum bias data for the correlation of  two hadrons with
$p_T$$>$0.1 GeV/c show a pattern that appears to be a linear
combination of the patterns for flux tube fragmentation and hard
scattering \cite{CMS09}.  Results in Fig. 2, together with the two
hadron correlation data from lower energies \cite{Mak15} and higher
energies \cite{CMS09} reveal that the $p_{Tb}$ boundary of separation
moves to lower $p_T$ values as the collision energy increases.

Similarly, because of local conservation of strangeness in $q\bar q$
production, adjacently produced hadrons with opposite strangeness are
correlated back-to-back in azimuthal angles.
\vspace*{-1.5cm}
\hspace{4.8cm}
\begin{figure}[H]
\hspace{-0.4cm}
\includegraphics[width = 270 pt, height = 170 pt]{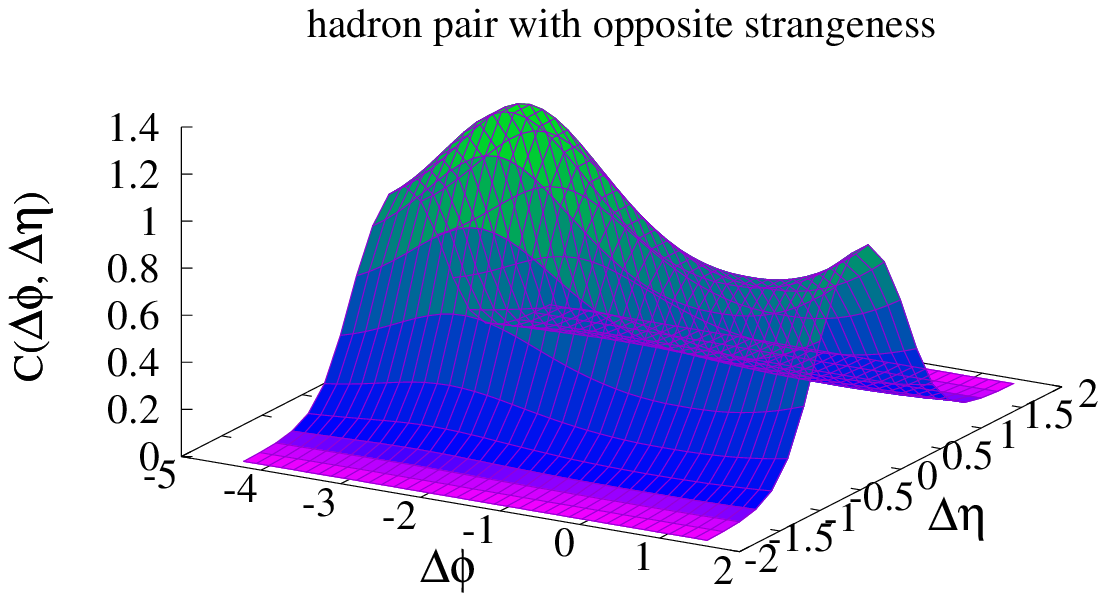}
\end{figure}

\vspace*{-5.400cm}
\hangafter=-11 \hangindent=3.6in 
\noindent Adjacently produced mesons however can be signaled by their
rapidity difference $\Delta y$ falling within the window of $|\Delta y
| $$\sim$$ 1/(dN/dy)$.  The theoretical correlation function for two
primary hadrons with opposite strangeness is shown in Fig. 3, with a
suppression at $(\Delta \phi, \Delta \eta)\sim$ 0 and an enhancement
at $\Delta \phi$$\sim$$ \pi$, within the window of $|\Delta y
|$$\sim$$ 1/(dN/dy)$.

\vspace*{0.2cm} 
\hangafter=-3 \hangindent=-7.4cm {\noindent{\bf
Fig.3}  The correlation function for two hadrons with opposite strangeness 
in  flux-tube fragmentation. }

\section{Conclusions and Discussions}

\hspace*{0.3cm} In the fragmentation of a flux tube, the production of
$q \bar q$ pairs obey local conservation laws and the hadrons follow
space-time-rapidity ordering.  As a consequence, a signature of the
flux tube fragmentation consists of a chain of produced hadrons,
correlated in rapidities, charges, flavors, and azimuthal angles.  The
observation of the chain of hadrons requires the identification of the produced
hadrons which may be possible if the fraction of unobserved hadrons is
small.

Another signature uses two-hadron angular correlations with opposite
charges or strangeness, which exhibits a suppression at $(\Delta
y,\Delta \phi)$$\sim$0, but an enhancement at $\Delta
\phi$$\sim$$\pi$, with the window of $|\Delta y |$$\sim$$ 1/(dN/dy)$.
It should be kept in mind however that resonance production and
resonance decay into hadrons will exhibit angular correlations similar
to the pattern of the flux-tube fragmentation.  The resonance fraction
give rise to complications and the two-hadron signature will work well
if the resonance fraction is not dominant.  Various estimates give the
resonance fractions to be of order 10 to 30\% for $pp$ collisions at
$\sqrt{s_{pp}}=17.3$ GeV \cite{Won15a}.

In flux-tube fragmentation, the production of two adjacent hadrons
with like charges or the same strangeness is prohibited \cite{Won15}.
Experimentally, two-hadron angular correlation of like charges with $\Delta \eta$$\sim$0 is not zero \cite{Por05,Mak15}
which indicates that there may be an additional mechanism for like
charge production with $( \Delta \phi, \Delta
\eta)$$\sim 0$.  Further theoretical search for the origin of the
source of like charge and strangeness correlations at $( \Delta \phi,
\Delta \eta)$$\sim 0$ will be of interest.

\vspace*{0.3cm}

\noindent {\bf Acknowledgments}
This work was supported in part by the Division of Nuclear
Physics, U.S. Department of Energy, under Contract No. DE-AC05-00OR22725.

\end{document}